\documentclass[twocolumn,showpacs,preprintnumbers,nofootinbib,prd,superscriptaddress,groupedaddress,10pt]{revtex4-1}
\usepackage{graphicx,amssymb,amsmath,amsthm,amsfonts,epsfig}
\usepackage[linktocpage]{hyperref}
\usepackage[usenames]{color}
\usepackage{epstopdf}

\usepackage{bm}
\usepackage{dcolumn}
\usepackage[latin1]{inputenc}
\usepackage{latexsym}
\usepackage{rotating}
\usepackage{hyperref}
\usepackage{color}
\usepackage{longtable}
\usepackage{enumerate}
\usepackage{mathtools}
\usepackage{url}
\def\p{\partial}
\def\na{\nabla}

\newcommand{\ben}{\begin{enumerate}}
\newcommand{\een}{\end{enumerate}}

\def\be{\begin{equation}}
\def\ee{\end{equation}}
\def\bea{\begin{eqnarray}}
\def\eea{\end{eqnarray}}
\newcommand{\beq}{\begin{eqnarray}}
\newcommand{\eeq}{\end{eqnarray}} 
\newcommand{\ba}{\begin{align}}
\newcommand{\ea}{\end{align}}

\def\eps{\epsilon}

\def\si{\sigma}

\def\ba{\bar{a}}

\begin{document}

\title{Electromagnetic emission from axionic clouds\\
and the quenching of superradiant instabilities}

\author{
Taishi Ikeda$^{1}$,
Richard Brito$^{2,3}$,
Vitor Cardoso$^{1,4}$
}
\affiliation{${^1}$ CENTRA, Departamento de F\'{\i}sica, Instituto Superior T\'ecnico -- IST, Universidade de Lisboa -- UL,
Avenida Rovisco Pais 1, 1049 Lisboa, Portugal}
\affiliation{${^2}$ Max Planck Institute for Gravitational Physics (Albert Einstein Institute), Am M\"{u}hlenberg 1, Potsdam-Golm, 14476, Germany}
\affiliation{${^3}$ Dipartimento di Fisica, ``Sapienza'' Universit\`a di Roma \& Sezione INFN Roma1, P.A. Moro 5, 00185, Roma, Italy}
\affiliation{${^4}$ Theoretical Physics Department, CERN 1 Esplanade des Particules, Geneva 23, CH-1211, Switzerland}

\begin{abstract}
The nature of dark matter is one of the longest-standing puzzles in science. Axions or axion-like particles are a key possibility, and arise in mechanisms to solve the strong CP problem but also in low-energy limits of string theory. Extensive experimental and observational efforts are actively looking for ``axionic'' imprints. Independently on their nature, their abundance, and on their contribution to the dark matter problem, axions form dense clouds around spinning black holes, grown by superradiant mechanisms.
It was recently suggested that once couplings to photons are considered, an exponential (quantum) stimulated emission of photons ensues at large enough axion number. Here we solve numerically the {\it classical} problem in different setups. We show that laser-like emission from clouds exists at the classical level, and we provide the first quantitative description of the problem.
\end{abstract}


\maketitle

\noindent{\bf{\em Introduction.}}
The existence of dark matter (DM) is established beyond any reasonable doubt. It manifests itself
clearly and inequivocally through gravitational interactions, but its nature and properties remain as elusive today as a century ago. Decades of searches hinging on interactions between DM and the Standard Model have all come back empty-handed. The advent of gravitational-wave (GW) astronomy promises to open a new chapter in our understanding of this hitherto invisible universe~\cite{Barack:2018yly}. The equivalence principle assures us that DM behaves gravitationally as any other matter, and points to gravity
as the key to unlock some of the mysteries of the missing matter.

The number of possible candidates is too large to enumerate here, and DM could be composed of a number of different particles.
An appealing candidate are axions or axion-like particles, first introduced to solve the strong CP problem in QCD~\cite{Peccei:1977hh}. These are strong candidates for cold DM~\cite{Bergstrom:2009ib,Fairbairn:2014zta,Marsh:2014qoa}. A plenitude of ultralight axion-like bosons might also arise from moduli compactification in string theory. In this ``axiverse'' scenario, a landscape of light axion-like fields can populate a mass range down to the cosmological scale, $\sim 10^{-33}\,{\rm eV}$~\cite{Arvanitaki:2009fg,Marsh:2015xka}. Irrespective of their origin, the seemingly empty arena between the electroweak (MeV) and the cosmological scale raises the question of weather new fields might exist somewhere between those scales.

Ultralight axions are expected to couple very weakly to ordinary matter, making their detection extremely challenging~\cite{Sigl:2017wfx}. However -- even with negligible initial abundance -- such fields trigger superradiant instabilities around massive, spinning black holes (BHs)~\cite{Detweiler:1980uk,Cardoso:2005vk,Dolan:2007mj,East:2017ovw,East:2018glu,Brito:2015oca}. The instability extracts rotational energy away from the spinning BH and deposits it into an axion cloud with high occupation number~\cite{Brito:2015oca}. Eventually, GW-emission dominates over the superradiant growth, leading to a secular spin-down and decay of the cloud. Such systems are a promising source of GWs that can be detected with current and future detectors~\cite{Arvanitaki:2010sy,Arvanitaki:2014wva,Brito:2014wla,Arvanitaki:2016qwi,Baryakhtar:2017ngi,Brito:2017wnc,Brito:2017zvb,Hannuksela:2018izj}.

The above picture neglects the coupling to matter, expected to be very weak. However, the (quantum-) stimulated emission of photons can be enhanced in highly dense axionic environments~\cite{Kephart:1986vc,Kephart:1994uy}, leading to the conjecture that blasts of light
could be emitted from BH systems~\cite{Rosa:2017ury,Sen:2018cjt}.  All the studies so far relied on adiabatic and flat-space approximations, and some of the conclusions are contradictory. In this Letter (further details are available in Ref.~\cite{Boskovic:2018lkj}) we show, by numerically solving Maxwell's field equations coupled to an axion field in a Kerr background, that electromagnetic (EM) fields can be exponentially amplified in such environments, even at classical level. Our results provide convincing evidence that, for a given axion coupling, bursts of EM radiation are emitted by the cloud above a critical value for the axion's amplitude, with potentially observable consequences.

\noindent{\bf{\em Setup.}}
Our starting point is the action describing a real massive (pseudo)scalar field $\Phi$ with axionic couplings to the EM field (we use geometrical units $G=c=1$ unless otherwise stated),
\beq\label{eq:MFaction}
{\cal L}&=&\frac{R}{k}- \frac{1}{4} F^{\mu\nu} F_{\mu\nu} - \frac{1}{2} g^{\mu\nu} \p_{\mu}\Phi\p_{\nu} \Phi
        - \frac{\mu^{2}}{2} \Phi^{2} \nonumber \\
&-& \frac{k_{\rm axion}}{2} \Phi \,^{\ast}F^{\mu\nu} F_{\mu\nu}\,.
\eeq
The mass of the scalar $\Phi$ is given by $m_{\rm S} = \mu \hbar$, $F_{\mu\nu} \equiv
\na_{\mu}A_{\nu} - \na_{\nu} A_{\mu}$ is the Maxwell tensor and $\,^{\ast}F^{\mu\nu} \equiv \frac{1}{2}\eps^{\mu\nu\rho\si}F_{\rho\si}$
is its dual. We use the definition $\epsilon^{\mu\nu\rho\si}\equiv \frac{1}{\sqrt{-g}}E^{\mu\nu\rho\si}$ where $E^{\mu\nu\rho\si}$ is
the totally anti-symmetric Levi-Civita symbol with $E^{0123}=1$. The quantity $k_{\rm axion}$ is a constant.
The scalar and EM field satisfy the following equations of motion:
%
\begin{subequations}
\label{eq:MFEoMgen}
\begin{align}
\label{eq:MFEoMScalar}
&\left(\nabla^{\mu}\nabla_{\mu} - \mu^{2} \right) \Phi =\frac{k_{\rm axion}}{2} \,^{\ast}F^{\mu\nu} F_{\mu\nu}
\,,\\ 
\label{eq:MFEoMVector}
&\nabla^{\nu}F_{\mu\nu} = - 2 k_{\rm axion} \,^{\ast}F_{\mu\nu} \nabla^{\nu} \Phi\,.
\end{align}
\end{subequations}
%
While for generic axion-like particles $k_{\rm axion}$ is independent on the axion mass $m_{\rm S}$, for the QCD axion they are related by (see e.g.~\cite{Cheng:1995fd}),
\be
\frac{\sqrt{\hbar}}{k_{\rm axion}}\sim \left(10^{13}~\text{--}~10^{16}\right)\left(\frac{10^{-5}\,{\rm eV}}{m_{\rm S}}\right)\,{\rm GeV}\,,
\ee
where the specific order of magnitude is model-dependent. To be as general as possible, we instead take $k_{\rm axion}$ to be an additional free parameter of the theory. It can be shown that the backreaction of the axion and vector field on the geometry is negligible~\cite{Brito:2014wla}. Therefore, we focus on Eqs.\eqref{eq:MFEoMScalar} and \eqref{eq:MFEoMVector} on a fixed Kerr background.
We impose the Lorenz condition on the vector field,
\begin{eqnarray}
\nabla_{\mu}A^{\mu}=0.
\end{eqnarray}

To formulate Eqs.(\ref{eq:MFEoMScalar}) and (\ref{eq:MFEoMVector}) as a Cauchy problem,
we use the standard 3+1 decomposition of the metric,
\begin{eqnarray}
ds^2=-\alpha^{2}dt^{2}+\gamma_{ij}(dx^{i}+\beta^{i}dt)(dx^{j}+\beta^{j}dt)\,,
\end{eqnarray}
where $\alpha$ is the lapse function, $\beta^{i}$ is a shift vector, and $\gamma_{ij}$ is the 3-metric on spacial hyper-surface.
Furthermore, by using a normal vector $n^{\mu}$ to the spatial hyper-surface,
the vector field $A_{\mu}$ can be decomposed as
%
%
%
\be
A_{\phi}=-n^{\mu}A_{\mu}\,,\qquad \mathcal{A}_{i}=\gamma^{j}_{~i}A_{j}\,.
\ee
We also introduce the EM fields
\be
\displaystyle E^{i}=\gamma^{i}_{~j}F^{j\nu}n_{\nu}\,,\quad \displaystyle B^{i}=\gamma^{i}_{~j}~^{\ast}F^{j\nu}n_{\nu}\,,
\ee
and the scalar momentum $\Pi$,
\be
\Pi=-n^{\mu}\nabla_{\mu}\Phi\,.
\ee

Finally, we use the constraint damping variable $Z$ to stabilize the numerical time evolution. The evolution equation for the axion field is written as
\beq
\partial_{t}\Phi&=&-\alpha\Pi+\mathcal{L}_{\beta}\Phi\,,\nonumber \\
\partial_{t}\Pi&=&\alpha(-D^{2}\Phi+\mu^{2}\Phi+K\Pi-2k_{\rm axion}E^{i}B_{i})\nonumber\\
&-&D^{i}\alpha D_{i}\Phi+\mathcal{L}_{\beta}\Pi\,.\nonumber
\eeq
From Maxwell Eq.~\ref{eq:MFEoMVector} and the Lorentz gauge condition, we find
\beq
\partial_{t}\mathcal{A}_{i}&=&-\alpha(E_{i}+D_{i}\mathcal{A}_{\phi})-A_{\phi}D_{i}\alpha+\mathcal{L}_{\beta}\mathcal{A}_{i}\,,\\
\partial_{t}E^{i}&=&\alpha(KE^{i}+D^{i}Z-(D^{2}\mathcal{A}^{i}-D_{k}D^{i}\mathcal{A}^{k}))\nonumber\\
&+&2\alpha k_{\rm axion}(\epsilon^{ijk}E_{k}D_{j}\Phi+B^{i}\Pi)+\epsilon^{ijk}D_{k}\alpha B_{j}+\mathcal{L}_{\beta}E^{i},\nonumber\\
\partial_{t}A_{\phi}&=&\alpha(KA_{\phi}-D_{i}\mathcal{A}^{i}-Z)-\mathcal{A}_{j}D^{j}\alpha+\mathcal{L}_{\beta}A_{\phi}\,,\\
\partial_{t}Z&=&\alpha(D_{i}E^{i}-\kappa Z)+2k_{\rm axion}\alpha B_{i}D^{i}\Phi+\mathcal{L}{_\beta}Z\,.
\eeq
Finally, we get Gauss's law as a constraint equation:
\be
\label{Eq:Gauss's law}
D_{i}E^{i}+2k_{\rm axion} B_{i}D^{i}\Phi=0\,.
\ee
We use Cartesian Kerr-Shild coordinates $(t,x,y,z)$~\cite{Witek:2012tr}.

The quantities extracted from our numerical simulation are multipolar components of the physical variables, $\Phi_{0},\,\Phi_{1},\,FF_{i}\equiv (F_{\mu\nu}F^{\mu\nu})_i, (T_{tr}^{\rm EM})_i$, with
\begin{eqnarray}
X_{0}(t,r)&:=&\int d\Omega X(t,r,\theta,\phi) Y_{00}(\theta,\phi)\,,\\
X_{1}(t,r)&:=&\int d\Omega X(t,r,\theta,\phi) Y_{R}(\theta,\phi)\,.
\end{eqnarray}
$T_{tr}^{\rm EM}$ is the $(t,r)$ component of energy momentum tensor of the EM field, $Y_{\ell m}(\theta,\phi)$ are spherical harmonics and $2Y_{R}(\theta,\phi)=Y_{1,1}+Y_{1,-1}$.

Without loss of generality, we use as initial data the following profile,
\be
\label{Eq.axion cloud initial data}
\Phi(t,r,\theta,\phi)=A_{0}rM\mu^{2}e^{-rM\mu^{2}/2}\cos(\phi -\omega_{\rm R}t)\sin\theta\,,
\ee
shown to yield a good approximation to the bound states around a BH of mass $M$ and angular momentum $J=Ma$~\cite{Yoshino:2013ofa,Brito:2014wla}.
Here $A_{0}$ is an arbitrary amplitude related to the mass in the axion cloud, and $\omega_{\rm R}\sim \mu$ is the bound-state frequency. We use the constraint-satisfying initial data for the EM field, $\mathcal{A}_{i}=A_{\phi}=E^{r}=E^{\theta}=0$, and
\be
E^{\phi}=E_{0}e^{-\left(\frac{r-r_{0}}{w}\right)^{2}}\Theta(\theta)\,,\label{Eq.electromagnetic field initial data}
\ee
where $E_{0}, w, r_{0}$ are arbitrary parameters describing the initial data. Here, $\Theta(\theta)$ determines the $\theta$-dependence of the initial data, the results below refer to $\Theta=1$~\cite{theta_note}.

The evolution equations were integrated using fourth-order spatial discretization and a Runge-Kutta method.

\noindent{\bf{\em Results I. Instability in flat-space.}}
%
\begin{figure}[tb]
\begin{center}
\begin{tabular}{c}
\epsfig{file=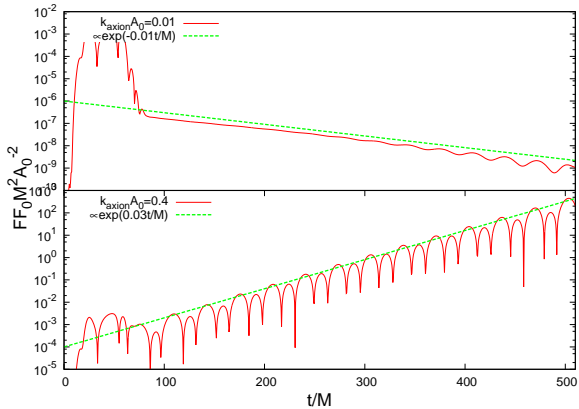,width=8.5cm,angle=0,clip=true}
\end{tabular}
\caption{
Time evolution of the monopole part of the EM scalar $F_{\mu\nu}F^{\mu\nu}$ for a coupling $M\mu=0.2$ at $r=20M$ when $k_{\rm axion}A_{0}=0.01$ (upper panel) and $k_{\rm axion}A_{0}=0.4$ (lower panel), in a Minskowski background. The scalar field is kept fixed and described by \eqref{Eq.axion cloud initial data}. The initial profile is described by $(E_{0}/A_{0},w/M,r_{0}/M)=(0.001,5.0,40.0)$, but the qualitative features of the evolution are independent on these. For small couplings $k_{\rm axion}A_0$ there is no instability and the initial EM fluctuation decays exponentially. For large couplings, on the other hand, an exponential growth ensues.
\label{graph_fixed_scalar}}
\end{center}
\end{figure}
A simple dimensional analysis indicates that the relevant quantity is $k_{\rm axion}\Phi$, independently on how the axion $\Phi$ was created or grown. This conclusion is consistent with a recent, flat-space analysis~\cite{Sen:2018cjt}. Therefore, to characterize the process we start with initial conditions \eqref{Eq.electromagnetic field initial data} for the EM field, while ``freezing'' the axion (i.e., the Klein Gordon equation is not evolved, and the axion is described by Eq.~\eqref{Eq.axion cloud initial data} at all times). Superradiant instabilities have a typically large timescale, and this should be a rather good approximation. The background spacetime is Minkowski. Both of these assumptions will be dropped when we discuss the Kerr background, where we confirm that these approximations provide a good qualitative picture of the problem.

We calculated the time evolution for several different $k_{\rm axion}A_{0}$, $\mu M=0.1,\,0.2,\,0.3$.
The behavior of the EM fields are shown in Fig.~\ref{graph_fixed_scalar}. These results are qualitatively the same for different backgrounds and different initial conditions. When the coupling $k_{\rm axion}A_{0}$ is small, the initial EM fluctuation dissipates and seems to vanish exponentially (a zero EM field is an exact solution of the field equations).

On the other hand, when the coupling $k_{\rm axion}A_{0}$ is larger than a certain critical value $k_{\rm axion}^{\rm critical}A_{0}$, the EM field grows exponentially, as is apparent from the bottom panel. We find that for supercritical couplings, the oscillation frequency of the electric or magnetic field is $\omega_{\rm EM}\sim \mu/2$, in agreement with previous works~\cite{Rosa:2017ury,Sen:2018cjt}.

\begin{figure}[tb]
\begin{center}
\begin{tabular}{c}
\epsfig{file=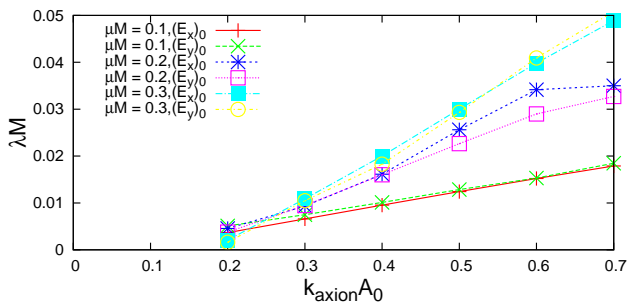,width=8.5cm,angle=0,clip=true}
\end{tabular}
\caption{The relation between the growth rate $M\lambda$ and coupling $k_{\rm axion}A_{0}$, in a Minskowski background with a scalar frozen to be given by \eqref{Eq.axion cloud initial data}. 
The rate is extracted from the $x$- or $y$-component of the electric field at $r=60M$. The instability is only present for large enough couplings, strongly suggesting the existence of a critical coupling $k_{\rm axion}A_{0}$. Such conclusions are consubstantiated by an analytical description of this system~\cite{Boskovic:2018lkj}.
\label{graph_lambda}}
\end{center}
\end{figure}
In the supercritical regime, the EM fields grow exponentially with time $\sim e^{\lambda t}$ (this agrees with one analytical analysis in flat space~\cite{Sen:2018cjt}, but not with the general statements on Kerr backgrounds~\cite{Rosa:2017ury}).
We estimate the exponential growth rate $\lambda$ of the electric field using best-fits to the local maxima.
The rate $\lambda$ (evaluated at $r=60M$) is shown in Fig.~\ref{graph_lambda} for different values of the coupling $k_{\rm axion}A_{0}$, and substantiates the claim that, even in Minkowski backgrounds there is indeed a critical coupling below which no instability sets in.
We also note that a flat space analysis of a uniform EM field and a uniform axion field finds
$\lambda_{0}=\mu k_{\rm axion}A_{0}$, with no critical coupling~\cite{Sen:2018cjt}. 
Our results for the rate are consistent with this estimate, in the supercritical regime. The critical value for the coupling is related to the non-uniformity of the field~\cite{Boskovic:2018lkj}. Since we are solely interested in axion clouds here, we do not explore this subject any further.

One important result born out of our numerical calculations is that a time-dependent axion is a crucial ingredient in the triggering of the instability (we did evolve also Maxwell equations for time-independent axions and no instability was observed). For $\Phi'(t)=$ constant, Eqs.~\eqref{eq:MFEoMVector} on a flat background can be reduced to a set of two-coupled radial differential equations for the vector components, which acquire an effective mass $\mu_{\rm eff}^2=-4k_{\rm axion}^2\Phi'(t)^2< 0$. For $\Phi'(r)=$constant, one finds $\mu_{\rm eff}^2=4 k_{\rm axion}^2\Phi'(r)^2> 0$, therefore suggesting that
the instability can be understood as a tachyonic-like instability, which is triggered only for time-dependent axions.

\noindent{\bf{\em Results II. Blasts of light in Kerr.}}
We now consider a background geometry described by a Kerr BH of mass $M$ and angular momentum $Ma^2$, and evolve both the axion and the EM fields. 

\begin{figure}[tb]
\begin{center}
\begin{tabular}{c}
\epsfig{file=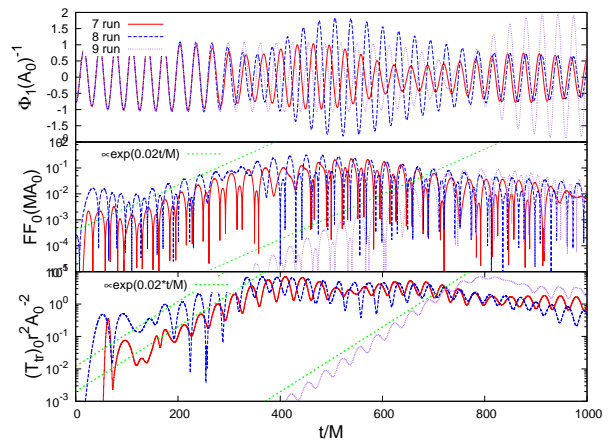,width=8.5cm,angle=0,clip=true}
\end{tabular}
\caption{
Time evolution of $\Phi_{1}$ (upper panel), $FF_{0}$ (middle panel) at $r=20M$ and the energy flux (bottom panel) for an axion with mass
$M\mu=0.2$ around a BH with $a=0.5M$. The coupling constant is super-critical with $k_{\rm axion}A_0=0.3$.
The initial EM profile is described by $(E_0/A_0,w/M)=(10^{-3},5),(10^{-3},20),(10^{-4},5)$ for run $7, 8, 9$ respectively, and $r_0=40$. The overall behavior and growth rate of the instability
at large timescales are insensitive to the initial conditions.
\label{graph_Kerr_sample}}
\end{center}
\end{figure}
We find that there is still a critical coupling beyond which an instability arises,
but now a new stage sets in at late times. This new stage appears because we evolve the axion as well. 
For super-critical values of the coupling $k_{\rm axion}$, the time evolution can be divided into three phases, as
seen in Fig.~\ref{graph_Kerr_sample}. The first, prompt phase, is a transient where the initial EM fluctuation travels through the axion cloud and dissipates.
During this phase, the EM field hardly affects the axion cloud. Subsequently, and excited by this prompt signal,
the axion cloud transfers (through the coupling) energy to the EM field, leading to a growth of the EM field and a suppression of the axion. We find that after this burst of radiation the new coupling is now below the critical value and the system is stable. A further unstable phase would need to wait for the depleted axion to be replenished via superradiance.
In line with the Minkowski results, the initial instability growth seems to be of the form $e^{\lambda t}$.

We find that the dipolar component of the EM field, $FF_1$, is nonzero and around two orders of magnitude smaller than $FF_0$ at the extraction radius if the figure.
When the EM fields becomes large enough, the energy dissipates as a burst of light.
The time evolution of the energy flux for a super-critical value of the coupling is shown in Fig.~\ref{graph_Kerr_sample}.
The flux indeed grows exponentially, at a rate which is independent of initial conditions (and which can be read off either from $FF_i$ or the fluxes).
Such exponential growth leads to energy dissipation, leading to axion depletion and to a final cloud which is no longer in the super-critical regime.

Our results are only weakly dependent on the BH spin, which controls the superradiant growth and dictates
the size to which the cloud grows. But the timescales of the bursts that we're studying are much smaller than the typical superradiant timescales. Therefore
the only quantity of relevance here is the magnitude of the scalar and therefore of the coupling, which dictates if an instability occurs or not.

We found that the maximum value of $T_{tr}M^{2}A_{0}^{-2}$ (at $r=100M$, already far from the axion cloud which peaks at $r\sim 50M$ for $M\mu=0.2$) is typically $10^{-4}-10^{-3}$.
Using the relation between the amplitude of the axion cloud $A_{0}$ and the mass of the cloud $M_{\rm S}$ \cite{Brito:2014wla},
we find the following peak luminosity for $\mu M=0.2$, and $k_{\rm axion}A_{0}=0.3-0.4$,
\begin{eqnarray}
\frac{dE}{dt}=5.0\times 10^{-6}\left(\frac{M_{\rm S}}{M}\right)\frac{c^{5}}{G}\,.
\end{eqnarray}
Our results show that for sufficiently large couplings $k_{\rm axion}$, axion clouds around BHs will eventually transfer a fraction of its energy to the EM field. Using a critical value for the instability $k_{\rm axion}A_{0}\sim 0.2-0.3$, the relation between the amplitude of the axion cloud $A_{0}$ and the mass of the cloud $M_{\rm S}$ \cite{Brito:2014wla} we find that the instability develops for axion couplings
\begin{equation}\label{crit_kaxion}
k_{\rm axion}\gtrsim 2\left(\frac{M}{M_{\rm S}}\right)^{1/2}\left(\mu M\right)^{-2}\,,
\end{equation}
which implies
\begin{equation}
\frac{\sqrt{\hbar}}{k_{\rm axion}}\lesssim 6\times 10^{18}\left(\frac{M_{\rm S}}{M}\right)^{1/2}\left(\mu M\right)^{2}\,{\rm GeV}\,.
\end{equation}

Finally we note that for scalars in a Kerr background the dominant-mode superradiant instability timescale is given by $t_{\rm inst}\sim 48M/((a/M)(M\mu)^9)$~\cite{Detweiler:1980uk,Dolan:2007mj}, while our results suggest that the timescale for the EM instability is given by $t_{\rm EM}\sim(\mu k_{\rm axion}A_{0})^{-1} \sim 10 M k_{\rm axion}^{-1}\left(M/M_{\rm S}\right)^{1/2}\left(\mu M\right)^{-3}$, where we used the relation between the amplitude of the axion cloud $A_{0}$ and the mass of the cloud $M_{\rm S}$~\cite{Brito:2014wla}. From Eq.~\eqref{crit_kaxion} it follows that whenever $k_{\rm axion}$ is sufficiently large for the instability to occur then $t_{\rm EM}\lesssim 5 M/(M\mu)$. Therefore the EM instability always has much shorter timescales than superradiance.

\noindent{\bf{\em Results III. Blasts vs leakage.}}
%
\begin{figure}[tb]
\begin{center}
\begin{tabular}{c}
\epsfig{file=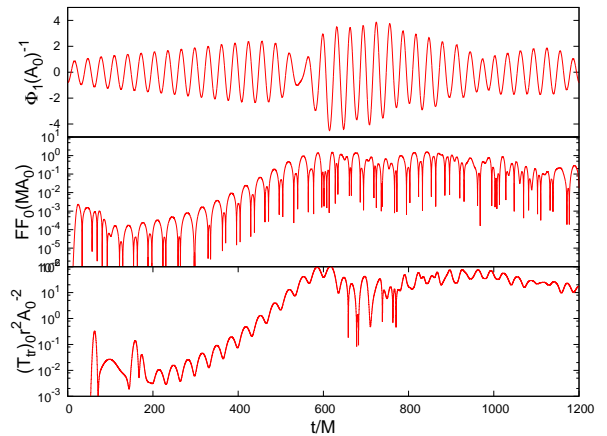,width=8.5cm,angle=0,clip=true}
\end{tabular}
\caption{
Evolution of an initially sub-critical axion, driven super-critical by a superradiant-like term. After the axion becomes super-critical, an instability sets in
which gives rise to a burst of EM radiation, leading to a depletion of the scalar, until a new superradiant growth set in. Our results are consistent with the triggering of 
periodic bursts. These results describe an axion with mass $M\mu=0.2$ around a BH with $a=0.5M$. The coupling constant is sub-critical $k_{\rm axion}A_0=0.15$.
The initial EM profile is described by $(E_0/A_0,w/M)=(10^{-3},5)$ and $r_0=40$.
\label{graph_Kerr_sample_C}}
\end{center}
\end{figure}
So far, we {\it assumed} that the axion grew through superradiance to some pre-determined value. If $k_{\rm axion}\Phi$ is supercritical at that point,
our results show that an instability kicks in and a EM blast ensues. A possible evolution of the system would then consist on stages of superradiant growth followed by EM blasts.
However, it is well possible that as the field grows, the evolution eventually leads to a constant EM flux locked into the slow superradiant evolution.
It is extremely challenging to test these two scenarios, since superradiant timescales are extremely large. However, one can introduce superradiant-like growth on shorter timescales
with the addition of a simple $C \partial\Phi/\partial t$ term to the Klein-Gordon equation~\eqref{eq:MFEoMScalar}. Such term was indeed used by Zel'dovich in his seminal study and can be shown to mimic accurately the correct description of many superradiant systems~\cite{zeldovich1,zeldovich2,Cardoso:2015zqa,Cardoso:2017kgn}. Such a Lorentz-invariance-violating term introduces a
superradiant-like instability, with a timescale of the order $1/C$, which we can tune to be within our numerical limits. A result of one such evolution with $C=4\times 10^{-4}$ is shown in Fig.~\ref{graph_Kerr_sample_C}. The axion is initially subcritical, but the superradiant-mimicking C-term drives it super-critical and triggers the instability. The instability proceeds as described previously. To conclude, our results show that bursts of EM radiation can indeed occur~\cite{GRIT}.

\noindent{\bf{\em Discussion.}}
The mere existence of light scalars will trigger superradiant instabilities around Kerr BHs and lead to
the depositing of the BH rotation energy in a ``cloud.'' This is a very generic feature, driven only by gravity.
Couplings to standard model field are expected to occur, and our results show that instabilities are triggered, wherein
a fraction of the cloud's energy is transferred to EM blasts.
The EM blasts carry a very precise frequency and can be described as laser-like emission. Our results
show a number of fine-details not present in previous simplified approaches to the quantum version~\cite{Rosa:2017ury,Sen:2018cjt}. For example, the EM field does {\it not} show signs of initial growth at the superradiant rate, nor do we see evidence for a $e^{t^2}$ burst. In addition, our results are consistent with the existence of a critical coupling, indicating that higher luminosities than previously reported may be possible.

For the ``typical'' coupling of the QCD axion with photons, this process is expected to only become relevant for axion masses above $\gtrsim 10^{-8}$ eV and therefore for BHs with masses $\lesssim 0.01M_{\odot}$, and could potentially explain fast radio bursts observed in the Universe. The smallness of the BH mass needed for this process to be efficient therefore implies that for the QCD axion it can only occur around hypothetical primordial BHs formed in the early Universe. However this limit is highly dependent on the coupling constant, and for generic axion-like particles it could become relevant for stellar mass BHs.

Our results have implications for the gravitational-wave detection of such systems~\cite{Arvanitaki:2010sy,Arvanitaki:2014wva,Brito:2014wla,Arvanitaki:2016qwi,Baryakhtar:2017ngi,Brito:2017wnc,Brito:2017zvb,Hannuksela:2018izj}, since EM bursts act as a limiter to the cloud size and therefore on the maximum amount of waves generated. Plasma effects and other observational issues are discussed in a separate publication~\cite{Boskovic:2018lkj} (see also Refs.~\cite{Sen:2018cjt,Yoshino:2015nsa}).

\begin{acknowledgments}
We thank Mateja Bo\v{s}kovi\'{c}, Ted Jacobson, Thomas Kephart, Masato Minamitsuji, Jo\~ao Rosa and Miguel Zilh\~ao for helpful comments.
The authors acknowledge financial support provided under the European Union's H2020 ERC 
Consolidator Grant ``Matter and strong-field gravity: New frontiers in Einstein's theory'' grant 
agreement no. MaGRaTh--646597. 
R.B. acknowledges financial support from the European Union's Horizon 2020 research and innovation programme under the Marie Sk\l odowska-Curie grant agreement No. 792862.
This project has received funding from the European Union's Horizon 2020 research and innovation programme under the Marie Sk\l odowska-Curie grant agreement No. 690904.
The authors would like to acknowledge networking support by the GWverse COST Action CA16104, ``Black holes, gravitational waves and fundamental physics.''
Computations were performed on the ``Baltasar Sete-Sois'' cluster at IST and XC40 at YITP in Kyoto University.
\end{acknowledgments}

\bibliographystyle{h-physrev4}
\bibliography{axion_refs}

\begin{thebibliography}{10}
\bibitem{theta_note}
{We also studied initial data described by $\Theta(\theta)=\sin^{4}(4\theta)$
  for $0\leq\theta<\frac{\pi}{4}$ and zero otherwise. The results are
  qualitatively the same as those described in the main text, confirming a
  flat-space analysis that shows that the instability is initial-data
  independent.}
\end{thebibliography}

\end{document}